\documentstyle[epsf,aps]{revtex}

\newcommand{\teb}[1]{\vec{#1}}
\newcommand{\bigint}{\displaystyle{\int}}
\newcommand{\bigfrac}[2]{\displaystyle{\frac{#1}{#2}}}
\begin{document}
\draft

\title{Finite temperature gluon self-energy in a class of temporal gauges}

\author{F. T. Brandt, J. Frenkel and F. R. Machado}
\address{Instituto de F\'\i sica,
Universidade de S\~ao Paulo\\
S\~ao Paulo, SP 05315-970, BRAZIL}

\date{\today}
\maketitle

\begin{abstract}
The approach which relates thermal Green functions to forward
scattering amplitudes of on-shell thermal particles is applied 
to the calculation of the gluon self-energy, in a class of 
temporal gauges. We show to all orders that, unlike the case of 
covariant gauges, the exact
self-energy of the gluon is transverse at finite temperature.
The leading $T^2$ and the sub-leading
${\rm ln}(T)$ contributions are obtained for temperatures $T$ high 
compared with the external momentum.
The logarithmic contributions have the same structure
as the ultraviolet pole terms which occur at zero
temperature.
\end{abstract}

\bigskip
\noindent
\section{Introduction}
Thermal gauge field theories in the temporal gauge has been the
subject of many investigations both in the imaginary and in the 
real time formalisms
\cite{kajantie:1985xx,kobes:1989up,james:1990it,james:1990fd,james:1991dz,leibbrandt:1994ki}. One of the main advantages of the 
{\it non-covariant} temporal gauge is that it is physical and
effectively ghost-free. At finite temperature, it may be considered
a more natural choice, since the Lorentz invariance is already
broken by the presence of the heat bath. It is also convenient
for calculating the response of the QCD plasma to a chromo-electric
field, since in this case only the gluon self-energy is 
needed (in this gauge, the chromo-electric field depends only linearly on the
gauge field $A^a_\mu$) \cite{kajantie:1985xx,kapusta:book89}.
Despite these advantages, explicit calculations of loop corrections
to Green functions are known to be more complicated than in
covariant gauges. One of the difficulty is associated with the
more involved tensorial structure of the propagator 
\hbox{(see Eq. (\ref{propag}))}. The other more fundamental 
problem is how to deal with the extra poles at $q\cdot n=0$ 
in the gluon propagator, where $q$ is the loop momentum and 
$n=(n_0,\vec 0)$ ($n_0^2>0$) is the {\it temporal axial} 
four-vector.

In the imaginary time formalism, the standard method of 
calculation of thermal Green functions consists in replacing
the Matsubara frequency sum by a {\it contour integral} in
the complex plane \cite{kapusta:book89,lebellac:book96}.
Using this approach, all Green functions, at one-loop order,
can be written as the sum of the {\it vacuum part} plus
the a {\it thermal part} which involves the Bose-Einstein 
(boson loop) or Fermi-Dirac (fermion loop) statistical 
distributions.

In the case of the temporal gauge, the extra poles 
at $q\cdot n=0$ in the gluon propagator demand more care
and make the standard {\it contour integral method} of
calculation very involved. In order to deal with this problem, 
Leibbrandt and Staley \cite{leibbrandt:1994ki}
have developed a technique for performing perturbative 
calculations at finite temperature in the temporal gauge
(see also reference \cite{james:1990it} for a different
approach to this problem). Employing a special
version of the {\it Mandelstam-Leibbrandt prescription}
\cite{mandelstam:1983cb,leibbrandt:1984pj} combined with the
{\it $\zeta$-function method} \cite{brandt:1991fs}, they
have obtained the {\it complete} $1/T$-expansion for the
one-loop self-energy component $\Pi_{00}^{ab}(k_0=0,\vec k)$. Their
results show that the leading and sub-leading contributions
to the self-energy can be unambiguously determined in the
temporal gauge, and are in agreement with the standard 
{\it contour integral} method employed in
reference \cite{kajantie:1985xx}.

The temperature dependent part of
Green functions can be described in terms of
{\it forward scattering amplitudes} of on-shell thermal
particles of the thermal medium. This idea was described
in reference \cite{barton:1990fk} and has been further
elaborated in the {\it Feynman gauge} and shown to be very
useful for determining the partition function in QCD at
high temperature \cite{frenkel:1990br}.
More recently this result has been generalized to the
case of {\it general covariant gauges} \cite{brandt:1997se}.
This method has been derived using both the imaginary time
formalism and the real time formalism up to two-loop order
\cite{brandt:1999gb,brandt:1999gf}. There is also an interesting
relation with the path integral approach \cite{brandt:1996fq}.

One of the purposes of this work is to extend the
forward scattering method to a class of 
temporal gauges. The main advantage of this method is that
in the high-temperature limit, it is straightforward 
to obtain the {\it full tensor structure} of both the leading 
$T^2$ and the sub-leading logarithmic contributions for 
temperatures $T$ such that $T\gg k$, where $k$ denotes the 
external four-momentum. For the leading $T^2$ part we obtain
the well known gauge invariant {\it hard thermal loop} result.
The {\it gauge dependent} sub-leading logarithmic part
shares with the previous calculations in general covariant 
gauges \cite{brandt:1997se} the interesting property of having 
the same structure as the {\it ultraviolet pole contributions} which 
occur at zero temperature. Another purpose of this work is to 
study 
the transversality property of the thermal
gluon self-energy. As is well known
\cite{kobes:1989up,brandt:1997se,weldon:1996kb}, in 
general covariant gauges, the exact self-energy of the gluon is not 
transverse at finite temperature. 

In section II we derive, in a class of 
temporal gauges, 
the full tensor structure of the forward scattering amplitude
associated with the gluon self-energy. In section III we show that, 
in contrast to the behavior in general covariant gauges, the exact thermal
self-energy of the gluon is transverse to all orders. In section IV we compute
the leading gauge-invariant $T^2$ and the sub-leading ${\rm ln}T$
contributions of the non-vanishing transverse structures.
The later is shown to be simply related to the ultraviolet pole
terms which occur at zero temperature. Finally, we present
a brief conclusion in section V.

\section{The forward scattering amplitude}
The diagrams which contribute to the gluon self-energy are shown
in the Figure 1. We employ the Feynman rules for the three- and 
four gluon vertices given in reference \cite{muta:book87}.
In the class of temporal gauges characterized by a gauge parameter 
$\alpha$, the gluon propagator can be written as 
\begin{equation}\label{propag}
\frac{1}{q^2}\left\{-i\delta^{ab}\left[g_{\mu\nu}-{1\over q\cdot u}
(q_{\mu}u_{\nu}+q_{\nu}u_{\mu})+{q_{\mu}q_{\nu}\over (q\cdot u)^{2}}
\left({\alpha\over n_{0}^{2}} q^{2}+1\right)\right]\right\} \, ,
\end{equation}
where the usual axial vector has been written as 
$n=n_0(1,\vec 0)\equiv n_0 u$.

\begin{figure}[h!]
    \epsfbox{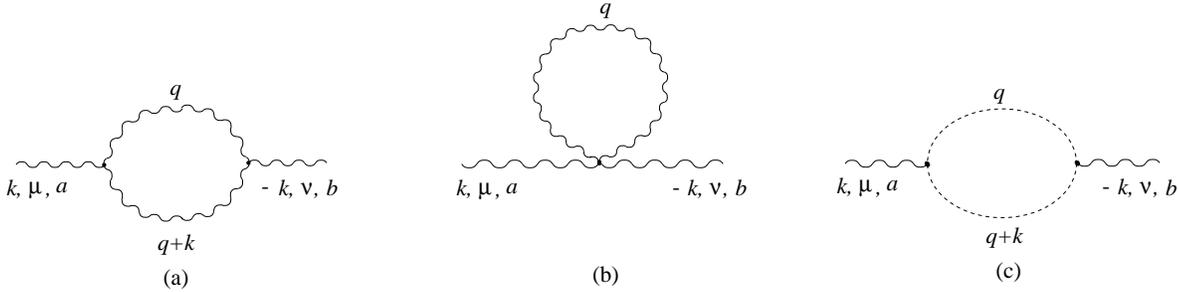}
\bigskip
   \label{fig1}
\caption{One-loop diagrams which contribute to the gluon-self
energy. Wavy and dashed lines denotes respectively gluons and
ghosts. All external momenta are inwards.}
  \end{figure}

Before considering the more involved diagrams in Figures (1a) and
(1b), let us show that the finite temperature contribution to the
diagram in Figure (1c) vanishes. In the imaginary time formalism
this contribution is proportional to the following 
integral \cite{kapusta:book89}
\begin{equation}\label{eq2}
\int d^{3}\vec q\int_{-i\infty+\epsilon}^{+\infty+\epsilon}
dq_{0}N(q_{0})\left[{t_{\mu\nu}\over 
n\cdot q\,n\cdot (q+k)}+q_0\leftrightarrow -q_0\right],
\end{equation}
where 
\begin{equation}\label{eq3}
N(q_0)=\frac{1}{{\rm e}^{\frac{q_0}{T}} - 1}
\end{equation}
is the Bose-Einstein statistical distribution
and $t_{\mu\nu}$ is a {\it momentum independent} tensor which
is obtained from the explicit expression for gluon-ghost vertex 
(note that the gluon-ghost vertex is not equal to zero 
in the class of gauges considered here). 
Using the partial fraction decomposition 
(for $n\cdot k$ different from zero)
\begin{equation}\label{eq4}
{1\over n\cdot q\,n\cdot (q+k)}=
{1\over n\cdot k}\left[{1\over n\cdot q}-{1\over n\cdot (q+k)}\right]
\end{equation}
and making the change of variable $q\rightarrow q-k$ in the
second term on the right hand side, 
we can see that Eq. (\ref{eq2}) {\it vanishes} when
one takes into account the property 
\begin{equation}\label{eq5}
N(q_0\pm k_0)= N(q_0\pm 2 n\pi i T)=N(q_0);
\;\; n=0,\pm 1,\pm 2 \cdots .
\end{equation}
This result holds independently of any explicit prescription 
for the poles at $q_0=0$.

Let us now consider the diagrams in Figures (1a) and (1b).
We are specifically interested in those contributions
which can be expressed in terms of forward scattering 
amplitudes of {\it on-shell} thermal particles. 
Following the steps described in the appendix \ref{appA},
such an expression can be obtained closing the contour of the 
$q_0$-integration in the right hand side of the complex plane 
and considering {\it only} the poles from the propagator 
(\ref{propag}) at $q_0=|\vec q|, |\vec q + \vec k| - k_0 $ 
and at $q_0=|\vec q|$, respectively for the diagrams in
the Figures (1a) and (1b). After shifting the momentum
in the contribution from the second pole in the diagram
of Figure (1a) and using the property (\ref{eq5}), we
obtain the following expression for the forward
scattering (FS) part of the gluon self-energy
\begin{equation}\label{eq6}
\left. \Pi^{ab}_{\mu\nu} \right |_{FS} =-\frac{1}{(2\pi)^3}
\int\frac{d^3\vec q N(|\vec q|)}{2|\vec q|}\frac 1 2
\left\{
  \begin{array}[h]{c}
\epsfbox{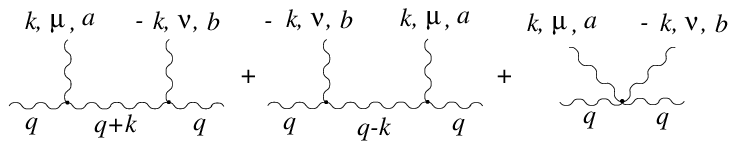}
  \end{array}
+q\leftrightarrow -q
\right\}_{q_0=|\vec q|},
\end{equation}
where the factor $1/2$ in front of the curly bracket takes
into account the symmetry of the graphs in figures (1a) and (1b).
It is understood that the Lorentz and color indices of the thermal 
particles (horizontal lines) are contracted with the tensor given 
by the curly bracket of Eq. (\ref{propag}). 

The full tensor structure generated by expression 
(\ref{eq6}) is rather lengthy. In order to express the
result in a more suggestive form, it is convenient to use
the following tensor decomposition \cite{weldon:1996kb}
\begin{equation}\label{eq7a}
\left.\Pi^{ab}_{\mu\nu} \right |_{FS}=
\delta^{ab}(\Pi_{T} P_{\mu\nu}^T + \Pi_{L} P_{\mu\nu}^L
+\Pi_{C} P_{\mu\nu}^C + \Pi_{D} P_{\mu\nu}^D),
\end{equation}
where
\begin{eqnarray}\label{eq7}
P_{\mu\nu}^T & = & g_{\mu\nu}-P_{\mu\nu}^L-
P_{\mu\nu}^D \, ,
\nonumber \\
P_{\mu\nu}^L & = & -\frac{\left(u\cdot k k_\mu -k^2 u_\mu\right)
                          \left(u\cdot k k_\nu -k^2 u_\nu\right)}
{k^2|\vec k|^2}\, , 
\nonumber \\
P_{\mu\nu}^C & = & {2k\cdot uk_{\mu}k_{\nu}-k^{2}(k_{\mu}u_{\nu}+k_{\nu}u_{\mu}) 
\over k^{2}|\vec k|} \, , \nonumber 
\nonumber \\ 
P_{\mu\nu}^D & = & {k_{\mu}k_{\nu} \over k^{2}} \, ,
\end{eqnarray}
is a set of four independent tensors which can be constructed 
using two Lorentz indices, the gluon four momentum $k$ and a vector
$u$ which describes the four velocity of the heat bath.
The tensors $P_{\mu\nu}^{T,L}$
are transverse with respect to the external four-momentum
$k^\mu$ and satisfy $k^i P_{i\nu}^{T}=0$ and 
$k^i P_{i\nu}^{L}\neq 0$ 
($i=1,2,3$), whereas the tensors $P_{\mu\nu}^{C}$ and
$P_{\mu\nu}^{D}$ are non-transverse.

Using the orthogonality of the tensors given in Eq. (\ref{eq7}) 
as well as the normalization 
$\frac 1 2 P_{\mu\nu}^T P^{\mu\nu\; T}=
P_{\mu\nu}^L P^{\mu\nu\; L}= 
-\frac 1 2 P_{\mu\nu}^C P^{\mu\nu\; C}=
P_{\mu\nu}^D P^{\mu\nu\; D}=1$,
it is straightforward to obtain the explicit expressions for 
$\Pi^L$, $\Pi^T$, $\Pi^C$, $\Pi^D$ from Eqs. (\ref{eq6}) and (\ref{eq7a}) 
(the tensor algebra was performed using the MapleVR3 version of the
symbolic computer package HIP \cite{hsieh:1992ti}).
The resulting expressions for $\Pi^C$ and $\Pi^D$ turn out to be
{\it equal to zero}.
The vanishing of the one-loop correction to $\Pi^D$ is a general 
consequence of the finite 
temperature Slavnov-Taylor identities \cite{weldon:1996kb} 
while the vanishing of $\Pi^C$ is a property which is verified by the 
forward scattering part of the gluon self-energy. From the
transversality property of $P_{\mu\nu}^L$ and $P_{\mu\nu}^T$ 
it follows that
\begin{equation}\label{eq8}
k^\mu \left. \Pi^{ab}_{\mu\nu} \right |_{FS} = 0
\end{equation}

\section{Transversality of the exact gluon self-energy}

In order to ascertain whether the transversality property of the 
thermal gluon self-energy holds beyond the approximation of forward 
amplitudes with on-shell particles, it is useful to study 
the consequences of the Becchi-Rouet-Stora identities 
\cite{becchi:1974md} on the 
structure of $\Pi_{\mu\nu}^{ab}$  at finite temperature. To derive 
these, we start from the effective action:
\begin{equation}\label{Teq1} 
I=\int d^{4}x d^{4}y J^{\mu a}(x)
\check G^{ab}_{\mu}(x-y)\eta^{b}(y)+
\frac 1 2 \int d^{4}x d^{4}y A^{\mu a}(x)\check{\Gamma}^{ab}_{\mu\nu}(x-y)
A^{\nu b}(y) + \cdots
\end{equation}
where the ellipsis denote 
ghost 
terms which are unimportant for our purpose \cite{muta:book87}.
$J^{\mu a}$ is the source of the BRS transformation, 
and $\eta^{b}$ is a ghost field (Although in 
our case the closed ghost loops do not contribute to 
Feynman diagrams \cite{frenkel:1976zk}, 
the ghost fields need not necessarily decouple from the BRS identity, which
involves open ghost lines). $\check G^{ab}_{\mu}$ is a quantity related to 
the gauge transformation of the gluon field $A^{\mu a}$, 
being given to lowest order by the covariant derivative 
$\delta^{ab}\partial^\mu - g f^{abc} A^{\mu c}$ \cite{muta:book87}.
The quadratic part $\check{\Gamma}^{ab}_{\mu\nu}$ is the sum of free 
kinetic energy, without the gauge fixing term and the 
self-energy:
\begin{equation}\label{Teq2} 
\check{\Gamma}^{ab}_{\mu\nu}(k_{0},\vec{k})=\delta^{ab}
(k_{\mu}k_{\nu}- k^2 g_{\mu\nu})+ \Pi^{ab}_{\mu\nu}(k_{0},\vec{k})
\end{equation}

The relevant BRS invariance of the gluon self-energy can now be 
written as 
\begin{equation}\label{Teq3}
\int d^{4}x {\delta I\over \delta A^{c}_{\mu}(x) }{\delta I\over 
\delta J^{\mu c}(x)}=0 .
\end{equation}

Differentiating this equation with respect to $A^{b}_\nu (y)$ 
and $\eta^{a}(z)$,  and setting the sources and fields equal to zero
leads in momentum space to the identity:
\begin{equation}\label{Teq4}
\check G^{\mu,ca}(k_0,\vec k)\check{\Gamma}^{cb}_{\mu\nu}(k_0,\vec k)=0
\end{equation}
Since, to lowest order, $\check G^{\mu,ca}$ is proportional to $k^{\mu}$, 
it follows from Eq. (\ref{Teq2}) that the above identity 
implies the relation:
\begin{equation}\label{Teq5} 
k^{\mu}\Pi^{ab}_{\mu\nu}= (k^{2}g_{\mu\nu}-k_{\mu}k_{\nu})
G^{\mu,ab} - G^{\mu,\, ca}\Pi_{\mu\nu}^{cb},
\end{equation}
where $G^\mu$ denotes the contributions of Feynman loops to
$\check G^\mu$.

Let us first consider the one loop order function $G^{\mu,ab}_{(1)}$,
which may be represented as shown in figure 2a.
\begin{figure}[h!]
    \epsfbox{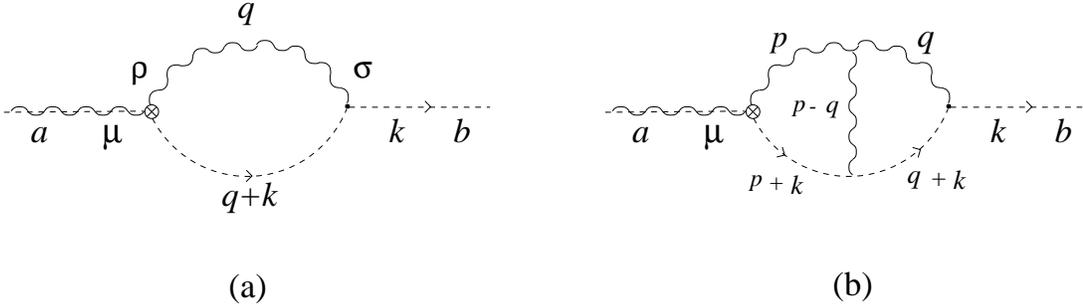}
\bigskip
   \label{fig2}
\caption{Examples of source-ghost diagrams.  
The traced/wave-line represents the external source and the crossed 
vertex is proportional to $g^{\mu\rho}$.}
  \end{figure}

Using Eq. (\ref{propag}) for gluon 
propagator and the fact that the gluon-ghost vertex is 
proportional to $n^{\sigma}$, one finds that $G^{\mu,ab}_{(1)}$ 
is given at finite temperature by
\begin{equation}\label{Teq6}
G^{\mu,ab}_{(1)}=\alpha g^{2}N \delta^{ab}(iT) 
\sum^{+\infty}_{l=-\infty} \int {d^{3-\epsilon}q\over 
(2\pi)^{3-\epsilon} } {q^{\mu}\over (q\cdot n)[(q+k)\cdot n]}
\end{equation} 
where $N$ is the number of colors, the space-time dimension is 
$4-\epsilon$ and $q_{0}=2\pi l i T$. 
One can now employ the identity (\ref{eq4})
and make the change of variable $q\longrightarrow q-k$
in the second term, to get
\begin{equation}\label{Teq7}
G^{\mu,ab}_{(1)}=\alpha g^{2}N \delta^{ab}{k^{\mu}\over k\cdot n}
(iT) \sum^{+\infty}_{l=-\infty} \int {d^{3-\epsilon}q\over 
(2\pi)^{3-\epsilon} } {1\over (q\cdot n)}
\end{equation}
Since this expression is 
orthogonal to $(k^{2}g_{\mu\nu}-k_{\mu}k_{\nu})$, it follows from
Eq. (\ref{Teq5}), that the exact gluon self-energy is transverse to
one-loop order.

In order to study the behavior of $G^{\mu, ab}$ in higher orders,
let us consider, for example, the two-loops diagram shown in 
Figure 2b. 
This yields at finite temperature a contribution given by:
\begin{eqnarray}\label{eq.1}
G^{\mu,ab}_{(2^\prime)}&=&(\alpha g^{2}N)^{2}\delta^{ab}(iT)^{2}
\sum^{+\infty}_{j=-\infty}\int {d^{3-\epsilon}p \over 
(2\pi)^{3-\epsilon}} {1\over (p+k)\cdot n}\left[ g^{\mu\sigma}
-{p^{\mu}n^{\sigma}\over p\cdot n}\right] \\ \nonumber
&&\times\sum^{+\infty}_{l=-\infty}\int {d^{3-\epsilon}q \over 
(2\pi)^{3-\epsilon}} {q_{\sigma}\over (q\cdot n)[(q+k)\cdot n]
[(p-q)\cdot n]}
\end{eqnarray}
where $p_{0}=(2\pi iT)j$ and $q_{0}=(2\pi iT)l$.

Using partial fraction decompositions and making appropriate changes
of variables, one finds that  Eq. (\ref{eq.1}) can be written 
in the form:
\begin{eqnarray}\label{eq.2}
G^{\mu,ab}_{(2^\prime)}&=&(\alpha g^{2}N)^{2}\delta^{ab}(iT)^{2}
\sum^{+\infty}_{j=-\infty}\int {d^{3-\epsilon}p \over 
(2\pi)^{3-\epsilon}} {1\over (p\cdot n)^{2}}\left[ {k^{\mu} 
\over k\cdot n}-{p^{\mu}\over p\cdot n}\right] \\ \nonumber
&&\times \sum^{+\infty}_{l=-\infty}\int {d^{3-\epsilon}q \over 
(2\pi)^{3-\epsilon}} {1\over (q\cdot n)}
\end{eqnarray}
which is rather typical of the contributions associated with two-loop diagrams.

Comparing the relation (\ref{eq.2}) with the one-loop result 
given  by Eq. (\ref{Teq7}), one can see that both expressions have the 
same $q$-dependence. But such contributions vanish, in 
consequence of the antisymmetry of $(q\cdot n)^{-1}$. Note 
that, thus far, we have made no assumptions about the 
specific form of $n$, so that this conclusion is also 
true in a 
class of axial gauges \cite{frenkel:1976zk}.

In the specific case of temporal gauges where $q\cdot n=q_{0}n_{0}$,it is
worth noticing that the integrals decouple, in fact, from the sums. Then,
we can set,in the sense of dimensional regularisation:
\begin{equation}\label{eq.3}
\int d^{3-\epsilon}q |\vec{q}|^{\omega}=0,\;\;\;\;\;\;
\omega\ge 0
\end{equation}
Furthermore, using the Mandelstam-Leibbrandt prescription to 
remove the singular terms at $p_{0}=0$ and $q_{0}=0$, the sum 
over $l$ also gives zero because of the antisymmetry of 
$q_{0}=(2\pi iT)l$.

The above arguments, which show that $G^{\mu,ab}_{(1)}$ and 
$G^{\mu,ab}_{(2)}$ are equal to zero, may be extended to higher order
to establish that $G^{\mu,ab}$ must generally vanish. It then follows
from  Eq. (\ref{Teq5}) that
the exact self-energy of the gluon is transverse at all temperatures. 
This important property implies that, to all orders, the structure functions 
$\Pi_{C}$ and $\Pi_{D}$ should be equal to zero.
(We show in the appendix \ref{appB} that these results 
are confirmed by explicit calculations to one-loop order.) In
consequence of this behavior, the non-linear relation among the
structure functions \cite{weldon:1996kb}
\begin{equation}\label{Teq8}
\Pi_{D}={\Pi_{C}^{2}\over k^{2}-\Pi_{L}}
\end{equation}
which must hold at finite temperature in any linear gauge, is
manifestly satisfied in the class of 
temporal gauges.

\section{High temperature contributions}
The non-vanishing structures $\Pi_{T}$ and $\Pi_{L}$, 
can be expanded as a series of high-temperature terms which 
come from the {\it hard thermal loop} region of {\it large} $q$,
such that $q \sim T >> k$.
(There are also other contributions, such as those proportional to $T$,
which originate from the
region of small $q$ \cite{kajantie:1985xx,leibbrandt:1994ki}.)
This can be done in a straightforward way expressing the
integrands in terms of the dimensionless four-vector 
$Q\equiv(1,\vec q/|\vec q|)$ and performing the expansion in 
powers of $1/|\vec q|$. The results up to sub-leading contributions 
are given by
\begin{eqnarray}\label{eq9}
&&\Pi_{T}^{htl}=-{g^{2}N \over 4(2\pi)^3|\vec{k}|^{2}}\int d\Omega
\left\{\left[-{4k^{4}\over (Q\cdot k)^{2}}+
               {8k^{2}k_{0}\over Q\cdot k}+
               4k^{2}-8k_{0}^{2}\right]
            \int d|\vec q|\,|\vec q|\,N(|\vec q|)
\right. \nonumber \\ 
&&+\left[-4Q\cdot kk^{2}k_{0}
        -{k^{8}\over (Q\cdot k)^{4}}
        +2{k^{6}k_{0}\over (Q\cdot k)^{3}}
        +4k^{2}k_{0}^{2}
        -8{k^{4}k_{0}^{2}\over (Q\cdot k)^{2}}
        +2k^{4} 
        +{5k^{6}\over (Q\cdot k)^{2}}
\right. \nonumber \\ 
&&
\left.\left.
        + {\alpha\over n_{0}^{2}}\left(     
         4 k^{4}k_{0}^{2}
        -2 k^{6}
        -4 Q\cdot k k^{4}k_{0}
        +2 (Q\cdot k)^{2}k^{4}  \right)
\right]
\int{d|\vec q|\over |\vec q|}N(|\vec q|)
\right\}
\end{eqnarray}
and
\begin{eqnarray}\label{eq10}
&&\Pi_{L}^{htl}=-{g^{2}N \over 4(2\pi)^3|\vec{k}|^{2}}\int d\Omega
\left\{ 
\left[{8k^{4}\over (Q\cdot k)^{2}}
      -{16k^{2}k_{0}\over Q\cdot k}
      +8k^{2}\right]
   \int d|\vec q|\,|\vec q|\,N(|\vec q|)
\right.
\nonumber \\ 
&&
+\left[8Q\cdot kk^{2}k_{0}
      -{4k^{6}k_{0}\over (Q\cdot k)^{3}}
      -16k^{2}k_{0}^{2}
      +{2k^{4}k_{0}^{2}\over (Q\cdot k)^{2}}
      +4k^{4}
      +{2k^{8}\over (Q\cdot k)^{4}} 
      +{4k^{6}\over (Q\cdot k)^{2}} 
\right. 
\nonumber \\ 
&&
\left.\left.
     -{\alpha\over n_{0}^{2}}  \left(
       4 k^{4}k_{0}^{2}
      +4 k^{2}(Q\cdot k)^{2}k_{0}^{2}
      -8 Q\cdot k k^{2}k_{0}^{3} \right)
      \right]
      \int{d|\vec q|\over |\vec q|}N(|\vec q|)\right\} 
\, ,
\end{eqnarray}
where $\int d\Omega$ denotes the integration over the directions
of $\vec q/|\vec q|$. Using that 
$\int d|\vec q| |\vec q| N(|\vec q|) = \pi^2 T^2/6$ and 
performing the angular integrations, we obtain
\begin{eqnarray}\label{eq11}
\Pi_{T}^{htl}=-g^{2}N\left\{ {T^{2}\over 12|\vec{k}|^{2}}\left[ 
{k^{2}k_{0} \over |\vec{k}|}
\ln \left(k_0+|\vec k|\over k_0-|\vec k| \right)-2k_{0}^{2}
\right]\right. \nonumber \\ 
-\left.{k^{2}\over 12\pi^{2}}\left(11-4{\alpha\over n_{0}^{2}}
k^{2}\right)
\int {d |\vec q|\over |\vec q|}N(|\vec q|) \right\}
\end{eqnarray}
\begin{eqnarray}\label{eq12}
\Pi_{L}^{htl}=g^{2}N\left\{ {T^{2}\over 6|\vec{k}|^{2}}\left[
{k_0 k^2 \over |\vec{k}|}
\ln \left(k_0+|\vec k|\over k_0-|\vec k|\right)-2 k^2\right]
\right. 
\nonumber \\
+\left.{k^{2}\over 12\pi^{2}}\left(11-4{\alpha\over n_{0}^{2}}
k_{0}^{2}\right)
\int{d |\vec q|\over |\vec q|}N(|\vec q|) \right\} \, .
\end{eqnarray}

The above contributions proportional to $T^2$ constitute the 
well known gauge independent hard thermal loop result 
which has been previously obtained in the Feynman gauge as well
as in general covariant gauges \cite{weldon:1982aq,brandt:1997se}.
It also agrees with the results obtained in the static 
and long-wavelength limits using the temporal gauge with
$\alpha=0$ \cite{kajantie:1985xx,leibbrandt:1994ki}.

From the second lines in Eqs. (\ref{eq11}) and (\ref{eq12}) 
one can extract the part proportional to ${\rm ln}{T}$
which comes from the region of high internal momenta. Such
contributions can be readily obtained noticing that  
$\int\frac{d |\vec q|}{|\vec q|}N(|\vec q|) = 
-\frac 1 2 {\rm ln}(T) + \cdots$. In this way, 
inserting the logarithmic part of 
Eqs. (\ref{eq11}) and (\ref{eq12}) into Eq. (\ref{eq7a}), 
we obtain the following result for the logarithmic contribution 
to the gluon self-energy in the temporal gauge 
\begin{eqnarray}\label{eq13}
\left.\Pi_{\mu\nu\,(\ln T)}^{ab}\right |_{FS}=
-{g^{2}N\delta^{ab}\over 8\pi^{2}}\left\{\left[
-{11\over3}+{4\alpha k^{2}\over 3n^{2}}\right](k_{\mu}k_{\nu}-k^{2}g_{\mu\nu})
\right. 
\nonumber \\
+\left.{4\alpha\over 3n^{4}}[(n\cdot k) k_{\mu}-k^{2}n_{\mu}][(n\cdot k)k_{\nu}
-k^{2}n_{\nu}]\right\}\ln T ,
\end{eqnarray}
where we have substituted $u=n/n_0$ and $n_0^2=n^2$.

Comparing the above logarithmic contribution with 
the dimensionally regularized zero-temperature gluon 
self-energy in the general axial gauge \cite{leibbrandt:1987ki}, 
we find that the coefficients of ${\rm ln}(T)$
and of $1/\epsilon$  are identical up to a relative sign. 

\section{Conclusion}
We have shown that, in a class of temporal gauges, the exact
thermal self-energy of the gluon is transverse to all orders.
(Except for the Feynman gauge, where the transversality has been verified 
only to one loop order,
this property is not valid in the 
class of covariant 
gauges\cite{weldon:1996kb}.) 
The forward scattering amplitude associated with the thermal
gluon self-energy has been obtained, to one-loop order, using the 
approximation which consists in neglecting the 
contributions from the {\it prescription poles}.
Our explicit calculations of the $T^2$ and $\ln(T)$ terms,
together with the arguments in \cite{leibbrandt:1994ki},
show that such corrections affect only 
the contributions proportional to powers of $1/T$, 
which become negligible at high temperature. Furthermore, our approach
preserves the transversality property of the exact thermal gluon self-energy.
The leading $T^2$ and the sub-leading ${\rm ln}(T)$ contributions are
consistently obtained from the forward scattering
amplitude of on-shell thermal particles. 
The relation between the dimensionally regularized zero
temperature gluon self-energy and the ${\rm ln}T$ contribution,
given in Eq. (\ref{eq13}), is relevant to explain 
the cancellation of the ${\rm ln}(-k^2)$ terms between the zero 
temperature and the temperature dependent parts of the gluon 
self-energy (such a cancellation has also been verified by explicit 
calculations in covariant gauges \cite{brandt:1999gm}).

\acknowledgements 
FTB and JF would like to thank CNPq (Brazil) for
financial support. FTB would like to thank the hospitality of the Department 
of Applied Mathematics of the University of Western Ontario (Canada)
where part of this work was completed and D. G. C. McKeon for helpful 
discussions. FRM would like to thank FAPESP for financial support. 

\appendix

\section{}
\label{appA}

In this appendix we illustrate the forward scattering amplitude technique
by considering the following type of integral which arises in the
calculation of the two-point function in the imaginary time 
formalism \cite{kapusta:book89}
\begin{equation}\label{genint}
I=\int{\rm d}^3\teb{q} \int_{-i\infty+\delta}^{+i\infty+\delta}
\frac{{\rm d} z}{2\pi i} N(z)
\left[\frac{t(q; p)}{\left(q\cdot q\right)^{i}
                               \left(p\cdot p\right)^{j}}
      + (z\leftrightarrow -z) \right],
\end{equation}
where $p\equiv q+k,\; k_0=2 m \pi i, m=0,\pm 1,\pm 2\cdots$, 
$q=(z,\teb{q})$, and we have factorized the denominators 
$1/(q\cdot q)^i$ from the propagators.
In covariant gauges $t(q; p)$ has no poles in the
complex $z$-plane, while for temporal gauges Eq. (\ref{propag})
yields extra {\it prescription poles}. For the ghost loop diagram in Fig 1c
we have (in the temporal gauge) the trivial situation $i=j=0$.
The two special cases $i=1,\,j=1$ and $i=1,\,j=0$ corresponds to Fig 1a and
Figs 1b, respectively. In what follows we will consider
the case $j\neq 0$ so that there will be contributions
from the poles at $q\cdot q =0$ and $p\cdot p =0$. The case
$j = 0$ can be easily derived using the same basic steps described below.

Since the integration in (\ref{genint}) is over all values of $\teb{q}$, 
it is convenient to make the change of variables 
$\teb{q}\leftrightarrow -\teb{q}$ in all the terms
$(z\leftrightarrow -z)$ so that 
\begin{equation}\label{genint2}
I=\int{\rm d}^3\teb{q} \int_{-i\infty+\delta}^{+i\infty+\delta}
\frac{{\rm d} z}{2\pi\,i} N(z)
\left[\frac{t(q; p)}{\left(q\cdot q\right)^{i}
                               \left(p\cdot p\right)^{j}}
      + q\leftrightarrow -q \right].
\end{equation}
Factorizing the denominators in (\ref{genint2}) we can write
\begin{equation}\label{genint3}
\begin{array}{ll}
I=\bigint{\rm d}^3\teb{q} \bigint_{-i\infty+\delta}^{+i\infty+\delta}
\bigfrac{{\rm d} z}{2\pi\,i} N(z) & \left[
\bigfrac{1}{\left(z+|\teb{q}|\right)^{i}}
\bigfrac{1}{\left(z+k_0+|\teb{p}|\right)^{j}} 
\right. 
\\ {}&{} \\
& \left.
\bigfrac{t(q; p)}{\left(z-|\teb{q}|\right)^{i}
\left(z+k_0-|\teb{p}|\right)^{j}}
+ q\leftrightarrow -q \right]
\end{array}.
\end{equation}
The $z$ integration can be readily performed using the Cauchy theorem
and closing the contour in the right hand side plane where
there are poles at $z=|\teb{q}|$ and $z=|\teb{p}|-k_0$
($k_0$ is a pure imaginary quantity at this stage of the calculation).
In this way we obtain
\begin{equation}\label{genint4}
\begin{array}{ll}
\left.I\right|_{FS}=-\bigint {\rm d}^3\teb{q} & 
\left\{\bigfrac{1}{({i}-1)!}
\lim_{q_0\rightarrow|\scriptsize{\teb{q}}|}\bigfrac{\partial^{{i}-1}}{
\partial q_0^{{i}-1}}\left(\bigfrac{N(q_0)}{(q_0+|\teb{q}|)^{i}}
                   \bigfrac{t(q;p)}{(p\cdot p)^{j}}\right) 
\right.
\\{}&{}\\
{}&\left. 
+ \bigfrac{1}{({j}-1)!}
\lim_{q_0\rightarrow|\scriptsize{\teb{p}}|-k_0}\bigfrac{\partial^{{j}-1}}{
\partial q_0^{{j}-1}}\left(\bigfrac{N(q_0)}{(q_0+k_0+|\teb{p}|)^{j}}
                   \bigfrac{t(q;p)}{(q\cdot
                     q)^{i}}\right)
+ q\leftrightarrow -q \right\}
\end{array},
\end{equation}
where the subscript $FS$ is to remind us that we have only considered the
{\it mass shell poles}.
Performing the change of variables $\teb{q}\rightarrow\teb{q} - \teb{k}$ in the
second term of (\ref{genint4}) we can write
\begin{equation}\label{genint5}
\begin{array}{ll}
\left.I\right|_{FS}=-\bigint {\rm d}^3\teb{q} & 
\left\{\bigfrac{1}{({i}-1)!}
\lim_{q_0\rightarrow|\scriptsize{\teb{q}}|}\bigfrac{\partial^{{i}-1}}{
\partial q_0^{{i}-1}}\left(\bigfrac{N(q_0)}{(q_0+|\teb{q}|)^{i}}
                   \bigfrac{t(q;p)}{(p\cdot p)^{j}}\right) 
\right.
\\{}&{}\\
{}&\left. 
+ \bigfrac{1}{({j}-1)!}
\lim_{q_0\rightarrow|\scriptsize{\teb{q}}|-k_0}\bigfrac{\partial^{{j}-1}}{
\partial q_0^{{j}-1}}\left(\bigfrac{N(q_0)}{(q_0+k_0+|\teb{q}|)^{j}}
                   \bigfrac{t(q_0,\teb{q}-\teb{k};q_0+k_0,\teb{q})}
                           {(q_0^2-|\teb{q}-\teb{k}|^2)^{i}}\right)
+ q\leftrightarrow -q \right\}
\end{array}.
\end{equation}
Finally, using the property $N(q_0\pm k_0)=N(q_0)$ and taking into 
account the contributions $q\leftrightarrow -q$ we obtain
\begin{equation}\label{genint6}
\begin{array}{ll}
\left.I\right|_{FS}=-\bigint {\rm d}^3\teb{q} & 
\left\{\bigfrac{1}{({i}-1)!}
\bigfrac{\partial^{{i}-1}}{
\partial q_0^{{i}-1}}\left(\bigfrac{N(q_0)}{(q_0+|\teb{q}|)^{i}}
                   \bigfrac{t(q;p)}{(p\cdot p)^{j}}\right) 
\right.
\\{}&{}\\
{}&\left. 
+ \bigfrac{1}{({j}-1)!}
\bigfrac{\partial^{{j}-1}}{
\partial q_0^{{j}-1}}\left(\bigfrac{N(q_0)}{(q_0+|\teb{q}|)^{j}}
                           \bigfrac{t(-p;q)}
                           {(p\cdot p)^{i}}\right)
+ q\leftrightarrow -q \right\}_{q_0 = |\vec q|}
\end{array}.
\end{equation}
Using $p=q+k$, the special case when ${i}={j}=1$ can be written as
\begin{equation}\label{genint7}
\left.I\right|_{FS}=-\int \frac{{\rm d}^3\teb{q}}{2\,|\teb{q}|}N(|\teb{q}|) 
\left[\frac{t(q;p)}{k^2+2 k\cdot q} + \frac{t(-p;-q)}{k^2-2 k\cdot q}
      + q\leftrightarrow -q \right]_{q_0 = |\vec q|}.
\end{equation}
The expression inside the bracket is a typical contribution 
to the on-shell forward scattering amplitude as represented, for example,
by the first two terms in the Eq. (\ref{eq6}).



\begin{thebibliography}{10}

\bibitem{kajantie:1985xx}
K. Kajantie and J. Kapusta, Ann. Phys. {\bf 160},  477  (1985).

\bibitem{kobes:1989up}
R. Kobes, G. Kunstatter, and K.~W. Mak, Z. Phys. {\bf C45},  129  (1989).

\bibitem{james:1990it}
K.~A. James and P.~V. Landshoff, Phys. Lett. {\bf B251},  167  (1990).

\bibitem{james:1990fd}
K.~A. James, Z. Phys. {\bf C48},  169  (1990).

\bibitem{james:1991dz}
K.~A. James, Z. Phys. {\bf C49},  115  (1991).

\bibitem{leibbrandt:1994ki}
G. Leibbrandt and M. Staley, Nucl. Phys. {\bf B428},  469  (1994).

\bibitem{kapusta:book89}
J.~I. Kapusta, {\em Finite Temperature Field Theory} (Cambridge University
  Press, Cambridge, England, 1989).

\bibitem{lebellac:book96}
M.~L. Bellac, {\em Thermal Field Theory} (Cambridge University Press,
  Cambridge, England, 1996).

\bibitem{mandelstam:1983cb}
S. Mandelstam, Nucl. Phys. {\bf B213},  149  (1983).

\bibitem{leibbrandt:1984pj}
G. Leibbrandt, Phys. Rev. {\bf D29},  1699  (1984).

\bibitem{brandt:1991fs}
F.~T. Brandt, J. Frenkel, and J.~C. Taylor, Phys. Rev. {\bf D44},  1801
  (1991).

\bibitem{barton:1990fk}
G. Barton, Ann. Phys. {\bf 200},  271  (1990).

\bibitem{frenkel:1990br}
J. Frenkel and J.~C. Taylor, Nucl. Phys. {\bf B334},  199  (1990).

\bibitem{brandt:1997se}
F.~T. Brandt and J. Frenkel, Phys. Rev. {\bf D56},  2453  (1997).

\bibitem{brandt:1999gb}
F.~T. Brandt, A. Das, J. Frenkel, and A.~J. da~Silva, Phys. Rev. {\bf D59},
  065004  (1999).

\bibitem{brandt:1999gf}
F.~T. Brandt, A. Das, and J. Frenkel, Phys. Rev. {\bf D60}, 105008 (1999)

\bibitem{brandt:1996fq}
F.~T. Brandt and D.~G.~C. McKeon, Phys. Rev. {\bf D54},  6435  (1996).

\bibitem{weldon:1996kb}
H.~A. Weldon, Annals Phys. {\bf 271},  141  (1999).

\bibitem{muta:book87}
T. Muta, {\em Foundations of Quantum Chromodynamics} (World Scientific,
  Singapore, 1987).

\bibitem{hsieh:1992ti}
A. Hsieh and E. Yehudai, Comput. Phys. {\bf 6},  253  (1992).

\bibitem{becchi:1974md}
C. Becchi, A. Rouet, and R. Stora, Commun. Math. Phys. {\bf 42},  127  (1975).

\bibitem{frenkel:1976zk}
J. Frenkel, Phys. Rev. {\bf D13},  2325  (1976).

\bibitem{weldon:1982aq}
H.~A. Weldon, Phys. Rev. {\bf D26},  1394  (1982).

\bibitem{leibbrandt:1987ki}
G. Leibbrandt, Rev. Mod. Phys. {\bf 59},  1067  (1987).

\bibitem{brandt:1999gm}
F.~T. Brandt and J. Frenkel, Phys. Rev. {\bf D60}, 107701 (1999).

\end{thebibliography}
%

\section{}
\label{appB} 

Here we show explicitly that the one-loop contributions to 
$\Pi_C$ and $\Pi_D$, in the class of temporal gauges
characterized by the gauge parameter $\alpha$,
are identical to zero. We start from the general expression
\begin{eqnarray}\label{app1}
\Pi_{C,D}&=&g^{2}N (iT) 
\sum^{+\infty}_{l=-\infty} \int {d^{3-\epsilon}q\over 
(2\pi)^{3-\epsilon} } {I}_{C,D}(k,q)
\nonumber \\
&=&\frac{g^2 N}{2\pi}\int{d^{3-\epsilon}q\over(2\pi)^{3-\epsilon}}
\int_{-i\infty+\delta}^{i\infty+\delta}{\rm d} q_0\frac 1 2
\left[I_{C,D}(k,q)+I_{C,D}(k,-q)\right]\coth\left(\frac{q_0}{2T}\right),
\end{eqnarray} 
where the integrands ${I}_{C,D}(k,q)$ are obtained projecting the
contributions from the diagrams in Figs. 1a and 1b onto the tensors 
$P_{\mu\nu}^{C,D}$. Using partial fraction decompositions, 
as in Eq. (\ref {eq4}), we perform shifts $q\rightarrow q - k$ 
in all the terms involving $(q+k)\cdot u$ in the denominator,
which are justified in the dimensional regularization scheme.
After using also the property 
$\coth((q_0\pm k_0)/2T)=\coth(q_0/2T \pm i\pi n)= \coth(q_0/2T)$
as well as the invariance of the $\vec q$-integral under 
$\vec q \rightarrow \vec q\pm \vec k$, the rather involved expressions 
of $I_{C,D}(k,q)$ simplify to
\begin{equation}\label{app2}
I_{C}(k,q)=\frac{1}{|\vec k|q^2(q+k)^2}\left[
8q\cdot u k\cdot q - k^2 k\cdot u - 6 k\cdot q k\cdot u -8(k\cdot q)^2
\frac{k\cdot u}{k^2} + 4 k^2 q\cdot u
\right]
\end{equation}
and
\begin{equation}\label{app3}
I_{D}(k,q)=\frac{1}{q^2(q+k)^2}\left[
2 k\cdot q - 4 q^2 + 8 \frac{(k\cdot q)^2}{k^2} - k^2
\right].
\end{equation}
It is interesting to note that the terms with denominators involving
factors like $q\cdot u$ or $(q+k)\cdot u$ cancel after performing
shifts, so that we do not need to worry about {\it prescription poles}. 
Hence the only poles which may contribute are the ones at
$q_0=|\vec q|$ and  $q_0=|\vec q + \vec k| - k_0$. The corresponding
residues of  $I_{C,D}(k,q)$ are 
\begin{equation}\label{app4}
{\rm Res}\left[I_C(k,q),q_0=|\vec q|\right]
=\frac{1}{2|\vec k||\vec q|}
\left.\left(4 q\cdot u - 
4 k\cdot u \frac{k\cdot q}{k^2} - k\cdot u 
\right)\right|_{q_0=|\vec q|},
\end{equation}
\begin{equation}\label{app5}
{\rm Res}\left[I_C(k,q),q_0=|\vec q+\vec k|-k_0\right]
=-\frac{1}{2|\vec k||\vec q + \vec k|}
\left.\left(4 q\cdot u - 
4 k\cdot u \frac{k\cdot q}{k^2} - k\cdot u 
\right)\right|_{q_0=|\vec q+\vec k|-k_0}
\end{equation}
\begin{equation}\label{app6}
{\rm Res}\left[I_D(k,q),q_0=|\vec q|\right]
=\frac{1}{2|\vec q|}
\left.\left(4 \frac{k\cdot q}{k^2} - 1
\right)\right|_{q_0=|\vec q|}
\end{equation}
and
\begin{equation}\label{app7}
{\rm Res}\left[I_D(k,q),q_0=|\vec q + \vec k|-k_0\right]
=-\frac{1}{2|\vec q+\vec k|}
\left.
\left(4 \frac{k\cdot q}{k^2} + 3
\right)\right|_{q_0=|\vec q+\vec k|-k_0}.
\end{equation}
Performing the shift $\vec q\rightarrow \vec q - \vec k$ 
in the Eqs. (\ref{app5}) and 
(\ref{app7}) one can see that the contributions from the
distinct poles in $I_{C,D}(k,q)$ exactly cancel each other.

\end{document}